\documentclass{iau} 
\usepackage{graphicx}

\title[Burst oscillations and dense matter] 
{Thermonuclear burst oscillations and the dense matter equation of state}

\author[Anna L. Watts]   
{Anna L. Watts}

\affiliation{Anton Pannekoek Institute for Astronomy, University of Amsterdam, Postbus 94249, 1090GE Amsterdam, the Netherlands \\ email: {\tt A.L.Watts@uva.nl} }

\pubyear{2017}
\volume{337}  
\jname{Pulsar Astrophysics -­ The Next 50 Years}
\editors{P. Weltevrede, B.B.P. Perera, L. Levin Preston \& S. Sanidas, eds.}

\begin{document}

\maketitle

\begin{abstract}
Matter in neutron star cores reaches extremely high densities, forming states of matter that cannot be generated in the laboratory.  The Equation of State (EOS) of the matter links to macroscopic observables, such as mass M and radius R, via the stellar structure equations. A promising technique for measuring M and R exploits hotspots
(burst oscillations) that form on the stellar surface when material accreted from a companion
star undergoes a thermonuclear explosion.  As the star rotates, the hotspot
gives rise to a pulsation, and relativistic effects encode information about M and R into the
pulse profile.  However the burst oscillation mechanism remains unknown, introducing uncertainty when inferring the EOS. I review the progress that we are making towards cracking this
long-standing problem, and establishing burst oscillations as a robust tool for measuring M
and R. This is a major goal for future large area X-ray telescopes.
\end{abstract}

\firstsection 
\section{Mysteries in dense matter}

Neutron stars offer a unique environment in which to develop and test theories of the strong force.
Densities in neutron star cores can reach up to ten times the density of a normal atomic nucleus, and
the stabilising effect of gravitational confinement permits long-timescale weak interactions. This
generates matter that is neutron-rich \cite[(Hebeler et al. 2015)]{Hebeler15}, and opens up the possibility of stable states of strange matter (either as deconfined quarks or in the form of hyperons), something that can only exist in neutron stars \cite[(see for example Chatterjee \& Vida\~na 2016)]{Chatterjee16}. 

Strong force physics is encoded in the Equation of State (EOS), the pressure-density relation, which links to macroscopic observables such as mass M and radius R via the stellar structure equations. By measuring and inverting the M-R relation we
can in principle recover the EOS and diagnose the underlying dense matter physics.  There are active efforts to do this using radio, X-ray and gravitational wave observations of neutron stars.  The techniques used vary, but all are based on the exploitation of relativistic effects. 

\section{Waveform modelling}

\begin{figure}[b]
\begin{center}
 \includegraphics[width=0.9\textwidth]{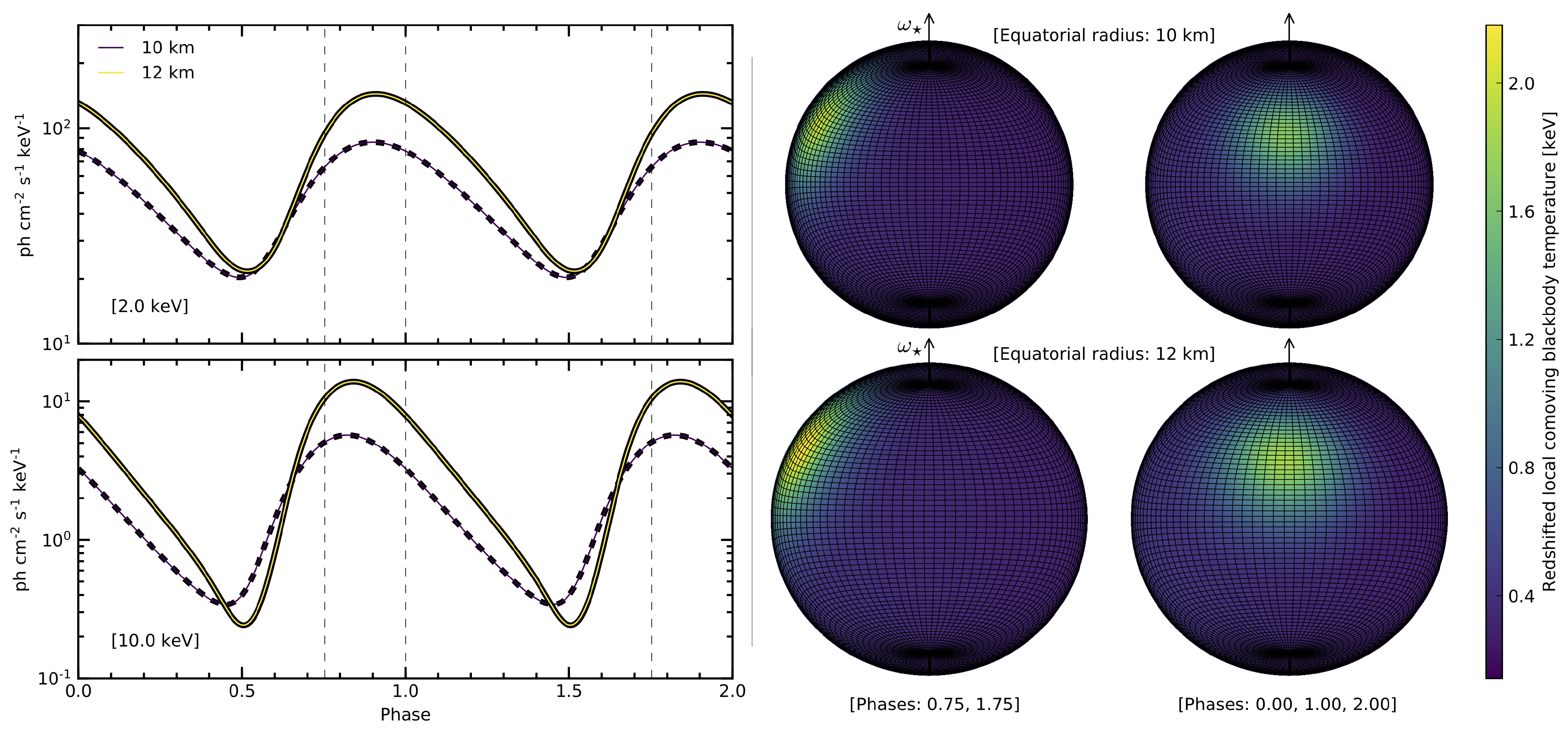} 
 \caption{EOS effects on monochromatic pulse profiles from a hotspot on a rotating neutron star, using a realistic space-time. The Figure shows surface patterns (right) and lightcurves (left) at two X-ray energies for different EOS models (with different equatorial radii) for a 1.8 $M_\odot$ neutron star rotating at 600 Hz.  Figure courtesy of Thomas Riley. }
   \label{fig2}
\end{center}
\end{figure}

One technique exploits the existence of hotspots (of various kinds) on the surfaces of neutron stars.  Modulated by the star's rotation, the hotspot give rise to a pulsation. Relativistic effects (Doppler boosting, gravitational redshifting, time delays and light-bending) encode information about M and R in the normalisation and harmonic content of the pulse profile (Figure \ref{fig2}).  By fitting the waveform, and marginalizing over nuisance parameters such as the surface pattern, one can recover information about M and R - and hence the EOS.  NICER is exploiting this technique to obtain EOS constraints for isolated X-ray pulsars (see contributions by P. Ray and S. Bogdanov to this volume).  This is possible because the space-time of rotating neutron stars is extremely well understood (see \cite[Watts et al. 2016]{Watts16}, and references therein).

\section{Thermonuclear burst oscillations}

\begin{figure}[b]
\begin{center}
 \includegraphics[width=0.9\textwidth]{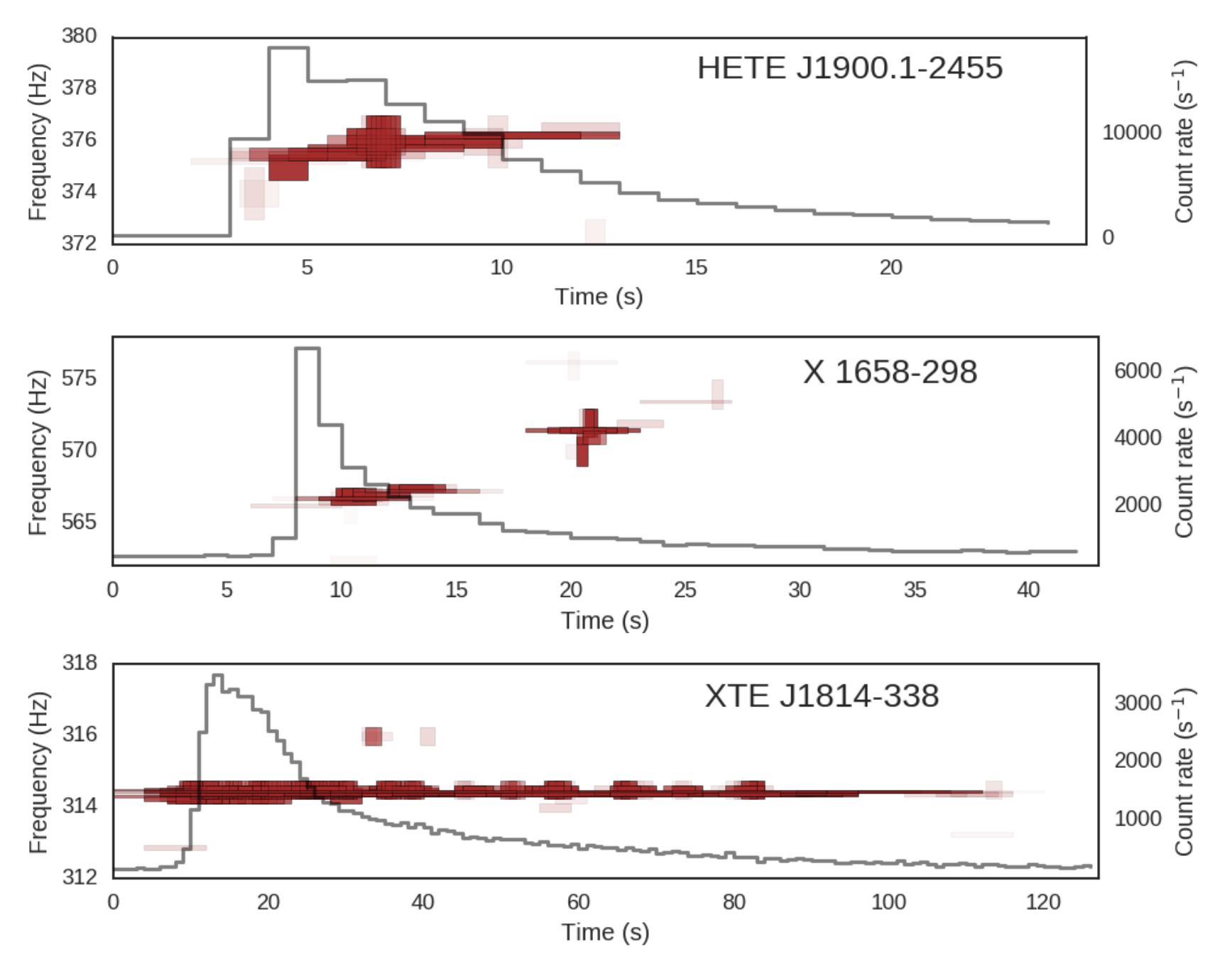} 
 \caption{Different burst oscillation behaviour from three sources, as recorded by the {\it Rossi 
X-ray Timing Explorer.} The plots show X-ray count-rate (thin black line, right axis) and the signal from several dynamical power spectra obtained with different Fourier transform lengths (frequency on left axis, transparency of contours indicating strength of signal). Strength and frequency of oscillation detections vary, as the surface pattern varies. Figure from Bilous et al. (in preparation).}
   \label{fig1}
\end{center}
\end{figure}

One class of hotspots to which the waveform modelling technique may be applied arise during Type I X-ray bursts: thermonuclear explosions in the oceans of accreting neutron stars, detectable in X-ray \cite[(for a review, see Strohmayer \& Bildsten 2006)]{Strohmayer06}).  The hotspots, known as burst oscillations, were discovered by \cite{Strohmayer96}, and have now been observed in bursts from 18 sources.  Figure \ref{fig1} shows some example burst oscillations: they have the following key properties (for reviews see \cite[Galloway et al. 2008]{Galloway08},  \cite[Watts 2012]{Watts12}):

\begin{enumerate}
\item{Frequencies are within a few Hz of the spin frequency (for sources where the spin is known independently), but can drift by up to a few Hz during a burst.}
\item{Amplitudes are in the range $\sim$ 20\% rms down to $\sim 5$\% rms (the detection threshold), with the highest amplitudes being reached during bursts at higher accretion rates.}
\item{For most sources, oscillations are only seen in some bursts; the exception being the bursting accretion-powered pulsars where (to date) all bursts show oscillations. }
\end{enumerate}
The fact that burst oscillations have not been detected in more sources seems to be a sampling artifact:  the highest amplitude signals are seen at higher accretion rates, and only a few sources have been observed extensively in the relevant accretion states with sufficiently sensitive instruments (\cite[Ootes et al. 2017]{Ootes17}).  This means that there are good prospects for new detections by current X-ray telescopes ASTROSAT and NICER.

Burst oscillations are particularly promising targets for waveform modelling for the following reasons:

\begin{enumerate}
\item{There are many known sources, making it more likely that we can map the M-R relation more completely.}
\item{The beaming pattern of the surface emission (thermal, through a scattering atmosphere) is well understood, hence not a source of modelling uncertainty.}
\item{Many of the sources exhibit other phenomena to which different EOS inference techniques can be applied (e.g. burst spectral modelling), meaning that multiple independent cross-checks are possible for the same source.}
\end{enumerate}
Large numbers of photons will still be required to deliver constraints at the required level of precision (\cite[Lo et al. 2013]{Lo13}); but this will be feasible with the next generation of large area X-ray telescopes (such as the proposed mission concepts eXTP and STROBE-X).   There then remains one major issue: the surface radiation pattern, which is an input to the waveform modelling process.  The burst oscillation mechanism is not yet known, and there are no a priori constraints on the pattern: moreover, as is clear from Figure \ref{fig1}, the pattern may evolve during a burst.    How then should we deal with this?  

\section{Resolving the outstanding issues}

\underline{What is the burst oscillation mechanism?}

Much uncertainty over the surface pattern and its evolution would be resolved if we could determine the nature of the burst oscillation mechanism (or mechanisms: the properties of burst oscillations from the pulsars are sufficiently different that these may well be caused by a different mechanism, see \cite[Watts 2012]{Watts12} and references therein).  The way that the thermonuclear flame spreads over the surface is important, and first principles simulations are now shedding light on the physics involved: conduction across an extended burning front, whose structure is controlled by rotational and magnetic field effects, is important (\cite[Cavecchi et al. 2013, 2015, 2016]{Cavecchi13,Cavecchi15,Cavecchi16}).  What happens after the flame has spread across the ocean is less clear.  Large-scale ocean modes, triggered by the burning (first suggested by \cite[Heyl 2004]{Heyl04}) still seem promising as a pattern generation mechanism, but the models would need to be modified to account for some key observational properties (see the discussion in \cite[Watts 2012]{Watts12}, and the contribution by F. Chambers to this volume).  We are also investigating the role of convection, expected in most bursts, which can give rise to zonal flows and associated patterns (see the contribution by F. Garcia to this volume). 

\underline{Characterizing the surface pattern uncertainty and its effects}

The {\it Rossi X-ray Timing Explorer} generated a rich archive of burst oscillation data. By combining that data with a relativistic ray-tracing code (such as that used to generate Figure \ref{fig2}) we can constrain the types of surface patterns (and their variability) that we observe.  By rigorously quantifying the level of uncertainty inherent in the surface pattern, and the levels of variation during bursts, we can assess the effects on the quality of EOS inference using this phenomenon.

\underline{Optimising observing and EOS inference strategy}

Obtaining sufficient photons for tight constraints on the EOS, using waveform modelling of burst oscillations, will likely require many observations with a total duration of hundreds of kiloseconds, even with a $\sim 10$m$^2$ telescope (\cite[Watts et al. 2016]{Watts16}).  The situation is complicated by the fact that many sources are transient, and will not be equally favourable in terms of the quality of EOS constraints that they deliver (due to e.g. geometry, location in M-R space).   Optimising this is now an active topic of study within the teams preparing the science case for future missions. 

\section{Summary}

It is 50 years since the discovery of neutron stars: but dense matter physics is not much older. Only 85 years have elapsed since the discovery of the neutron, and 53 years since the quark model made its appearance.  Thanks to X-ray astronomy (54 years old) we are making the first steps in exploiting a phenomenon that is a mere 21 years old (burst oscillations) to probe the dense matter in neutron star cores.  And the future?  No predictions! But there can be no doubt that it will be scientifically wonderful.
\\ 

{\bf Acknowledgements} This research is funded by ERC Starting Grant No. 639217 CSINEUTRONSTAR.

\end{document}